\documentclass[sigconf]{acmart}
\usepackage{booktabs}

\setcopyright{rightsretained}
\usepackage[]{soul}
\usepackage{dcolumn}
\newcolumntype{E}{>{\DDC}c}
\def\DDC#1(#2){\DC@..{3.2}#1\DC@end\,(\DC@..{2.1}#2\DC@end)}

\makeatother
\usepackage{siunitx}
\acmConference{10th International ACM Web Science Conference} 
\acmYear{2018}
\copyrightyear{2018}

\begin{document}

\copyrightyear{2018} 
\acmYear{2018} 
\setcopyright{acmcopyright}
\acmConference[WebSci '18]{10th ACM Conference on Web Science}{May 27--30, 2018}{Amsterdam, Netherlands}
\acmBooktitle{WebSci '18: 10th ACM Conference on Web Science, May 27--30, 2018, Amsterdam, Netherlands}
\acmPrice{15.00}
\acmDOI{10.1145/3201064.3201088}
\acmISBN{978-1-4503-5563-6/18/05}

\title{And Now for Something Completely Different:\\Visual Novelty in an Online Network of Designers}
 \author{Johannes Wachs}
 \affiliation{Central European University\\Budapest, Hungary}
 \email{wachs_johannes@phd.ceu.edu}
 \author{B\'alint Dar\'oczy}
 \affiliation{Institute for Computer Science and\\Control, Hungarian Academy of Sciences (MTA SZTAKI)\\Budapest, Hungary}
 \email{daroczyb@ilab.sztaki.hu}
 \author{Anik\'o Hann\'ak}
 \affiliation{Central European University\\Budapest, Hungary}
 \email{hannaka@ceu.edu}
  \author{Katinka P\'all}
 \affiliation{Institute for Computer Science and\\Control, Hungarian Academy of Sciences (MTA SZTAKI)\\Budapest, Hungary}
 \email{pall.katinka@sztaki.mta.hu}
  \author{Christoph Riedl}
 \affiliation{Northeastern University\\Boston, MA\\Harvard University\\Cambridge, MA}
 \email{c.riedl@neu.edu}
\renewcommand{\shortauthors}{J. Wachs, B. Dar\'oczy, A. Hann\'ak, K. P\'all, and C. Riedl}
\fancyhead{}

\begin{abstract}
Novelty is a key ingredient of innovation but quantifying it is difficult. This is especially true for visual work like graphic design. Using designs shared on an online social network of professional digital designers, we measure visual novelty using statistical learning methods to compare an image's features with those of images that have been created before. We then relate social network position to the novelty of the designer's images. We find that on this professional platform, users with dense local networks tend to produce more novel but generally less successful images, with important exceptions. Namely, users making novel images while embedded in cohesive local networks are more successful.
\end{abstract}



\maketitle

\section{Introduction}

High-quality creative design work can create tremendous value for organizations. It helps technical products gain acceptance~\cite{hargadonSutton1997} and it often serves as the basis for competition in cultural markets~\cite{wijnbergGemser2000}. Consequently, there has been mounting interest in the use of designers by organizations as a source of value creation~\cite{ravasiLojacono2005,rindovaDalpiazRavasi2011,rindovaPetkova2007}. One important ingredient to successful designs is novelty: the degree to which a design is new, original, or unusual relative to what has come before. One reason for this is that derivative work is frowned up in creative fields~\cite{bauer2016intellectual}. Indeed novelty is the prime ingredient of innovation and the production of new things~\cite{encinar2006novelty}. Economists have long known that innovation is the driving influence behind economic growth and development~\cite{schumpeter2003theory}, and recent studies suggest that successful companies make 80\% of their revenue with products younger than five years~\cite{kim1997value}. 

Despite its importance novelty is difficult to measure, especially in the context of creative design. In this paper we investigate three research questions related to novelty in design: (1) how can we measure novelty in digital design? (2) who produces novel work? and (3) what is the relationship between novelty and success? We develop and compare different mathematically-grounded measures of novelty or distinctiveness of digital images to better understand its antecedents and subsequent effect on success in a community of professional designers. 

To investigate these questions we collect roughly 40,000 images posted by over four thousand professional designers on an online community over a period of about four years. We propose and evaluate a measure of novelty for digital design at the image level using two feature sets: one capturing content and structure defined using an Inception neural network, the other capturing visual aesthetics using classical compositional features. We visualize the distributions of images in low dimensional projections of these feature spaces to better understand what these features capture and how they may capture novelty of an image.

We calculate novelty by comparing an image with prior images in terms of these derived features using information theoretic methods. This focus on temporal order distinguishes novelty from more ``timeless'' notions like beauty or appeal~\cite{elgammal2015quantifying}. Calculating novelty using the compositional features yields a measure of aesthetic or style novelty based on colors, spatial arrangement, and symmetry, while using Inception features results in a measure of content novelty. We validate our measures by showing that the earliest images annotated with emerging labels or ``tags" for new kinds of designs are indeed more content-novel.

With these measures of novelty for digital design in hand, we ask two questions: who produces novel images? How does novelty relate to success? The social networks literature makes two suggestions. Individuals with open, diverse social networks have access to diverse sources of information, which they may synthesize in novel ways~\cite{granovetter1973strength,burt2004structural}. But individuals in cohesive, closed networks have greater access to trust and social support, allowing them to more easily take the risk inherent in the creation of novelties~\cite{coleman1988social,krackhardt2003strength}. The literature suggests that when the domain is quickly changing and when the space of possible novelties is large, it is rather cohesive networks that facilitate novelty~\cite{aral2011diversity}. We argue that our topic of study is such a domain: design evolves quickly and new trends can be drastically different, and so we hypothesize that cohesive local networks do more to facilitate novelty in this domain than diverse ones.

Using a regression framework to analyze our panel data, we find a positive relationship between the local cohesion of a user's network on the site and the novelty of her images. Users in the global center of the network make less novel images. We suggest one possible explanation: that standing out is a form of risk-taking and that local network density facilitates this behavior. Furthermore, we find that novel images are on average less successful, but can be successful when originating from the right network position. Finally, we demonstrate that our novelty measures add explanatory power to a machine learning model predicting success, above and beyond a user's network position. This suggests that network position does not entirely mediate the relationship between novelty and success.

Our paper makes three contributions to the literature. First, we qualitatively compare the data encoded in different feature sets that can be derived from images. Second, we define a statistically-sound measure of the novelty of images, applicable to either set of features. Third, we provide empirical evidence for relationships between novelty, network position, and success, showing that novelty and network position together can predict success. 

\section{Related work}
In this section we first survey research on the quantification of novelty. Next, we overview literature on the relationship between novelty, social network position, and success, and finally, introduce studies that have looked at design in an online setting. 

\subsection{Quantifying Novelty}
As novelty is a complex construct with various dimensions, many different measures of it have been proposed~\cite{boudreau2016looking,fleming2001,redi2014}. One key notion underlying the measurement of novelty is the concept of recombination: that novelty is the result of reconfiguration of old ideas~\cite{weitzman1998}. Novelty is distinguished from aesthetic quality or beauty because it carries an intrinsic temporal property. For example, it is difficult to judge in retrospect how novel a product was at the time of its release. Previous studies on the beauty of images utilize the fact that crowdsourced judgments of beauty are relatively stable over time~\cite{schifanella2015image}. 

Recent models of novelty frame it in terms of the ``actual'' and the ``possible''. Models consider what it means for something to be new in terms of a path of discovery in an evolving complex space~\cite{loreto2016dynamics}. When something new is done for the first time, the space of the ``adjacent possible'' grows, making new things possible. In this framework, novelties are discrete, binary events. 

Previous work from the data mining community on novelty of images has mostly been concerned with the detection of outliers or anomalies within images, rather than across images~\cite{boracchi2014novelty,smola2009relative}. Most applications concern the detection of verifiable facts about an image: the presence of specific objects in satellite images, detecting biological abnormalities like cancer, etc. One commonality across these efforts, and indeed our own, is that features need to be extracted from an image to make computational analysis tractable.

Several recent data-driven studies quantify novelty in creative fields. In a study of popular music, Askin and Mauskapf compare songs with their predecessors using cosine similarity of a set of derived features like danceability and tempo~\cite{askin2017makes}. Past work has quantified the creativity of visual art as a combination of both novelty and influence using visual features~\cite{elgammal2015quantifying}. Redi et al. quantify the novelty of short video clips using a similar approach to ours~\cite{redi2014}, while Khosla et al. use image features to predict engagement on social media~\cite{khosla2014makes}. Natural language processing has also been applied to measure the novelty of textual content including scientific article abstracts~\cite{boudreau2016looking,evans2011metaknowledge} and equity crowdfunding campaigns~\cite{horvat2018crowd}. We do not define novelty of a thing in terms of success~\cite{sinatra2016quantifying} or surprise~\cite{barto2013novelty}. Novelty of a thing as we consider it says nothing intrinsically about its impact, influence, or outcomes. At the same time, we acknowledge that any attempt to measure novelty or distinctiveness can only capture a small facet of the phenomenon.

\subsection{Network Position, Novelty, Success}

Psychological research emphasizes that creativity is a demanding enterprise, requiring focus and concentration~\cite{csikszentmihalyi1996flow}. Given the apparent difficulty of creative endeavors, it is perhaps no surprise that social network structure plays a significant role in both facilitating novelty and shaping its reception. In fact, recent studies of creativity emphasize that novel products, even nominally created by a single author, can sometimes be understood as ``products of a momentary collective process'' ~\cite{hargadon2006collections}. How the networks that synthesize creative products fit together have strong predictive power of their eventual success~\cite{de2015game}.

So what kind of network position facilitates novelty? Creators embedded in a cohesive social network can hope to benefit from high amounts of social capital and support~\cite{coleman1988social}. Strong ties represent avenues of trust, which greatly facilitates the kind of risk-taking inherent in making a novel product in a professional, creative environment~\cite{krackhardt2003strength}. One study indicates that central actors in a network of research scientists produce more creative outputs, indicating that established actors can feel the freedom to experiment more broadly~\cite{perry2006social}. 

It is also true that diversity of social connections has been shown to foster creativity. Weak ties in social networks tend to bridge groups and provide an actor with access to novel information~\cite{granovetter1973strength}. Indeed the same study of research scientists cited above shows that creativity increases with the number of weak ties~\cite{perry2006social}. This line of thought is built on the idea that bridging actors occupying ``structural holes'' can create their own social capital by leveraging their unique access to diverse information~\cite{burt2004structural}. Whether open or closed networks better support novelty creation in our context is therefore an empirical question.

Besides the relationship of network position and novelty, the perception of novelty is also of interest to the research community. What ratio of traditional and novel maximizes success? Work across many disciplines find an inverse-U shaped relationship between novelty and success~\cite{boudreau2016looking,askin2017makes}. One prolific strand of the literature models novelty as the recombination of known ideas in new ways, and that the key to successful novelty is the combination of many conventional ingredients with relatively few new ones~\cite{uzzi2013atypical}.

\subsection{Online Design Communities}
Closest to our work empirically are studies on online design communities, like Dribbble, Behance, or Threadless. These studies generally focus on the question of how users or products become successful, and how different groups of users fare~\cite{deka2015ranking,riedl2018threadless}. For instance several studies find significant differences in the behavior and success of men and women on these sites~\cite{wachs2017men,kim2017creative}.

Dribbble has received attention from researchers because of its importance to the professional design community and its exclusive, invitation-only nature. In an interview-based study researchers found that users leverage the site and its social network to gather inspiration, learn skills by reverse engineering examples, anticipate trends in the marketplace, and to gather feedback~\cite{marlow2014rookie}. The study also found that users invested significant effort in developing a professional identity through the site. As in many other online communities, users reported the status importance of having many followers and collecting likes. 

More recently, machine vision researchers have taken an interest in learning from image data taken from online design communities, as they offer substantively different opportunities to develop machine vision than, say, photographs~\cite{wilber2017bam}. Similarly, the dual functions of online digital communities as places to post and places to be inspired offer interesting opportunities for bespoke recommendation systems~\cite{rudolph2016joint}.

\section{Data}
In this section we describe the Dribbble platform, our data collection method, and outline the extracted features at the image, user, and network levels.

\subsection{Dribbble}
Dribbble, founded in 2009, is an online community where designers share their work by posting images. It is a highly-visited site, with an Alexa rank of 1104, the second most popular website for design sharing after Behance.  Unlike most content-sharing platforms, the site operates on an invitation-only basis: though the site can be viewed by anyone, only invited users can post images. Active users are occasionally given invitations which they can use to invite other designers. Moreover, the number of images a user can post in a given time frame is capped. All together, this leads to high-quality content and the feeling of belonging to an ``elite'' community among users. 

The stakes on Dribbble are high. Interviews with users on the site reveal that individuals use the site to develop their professional identities~\cite{marlow2014rookie}. Indeed most users use their real names, post photographs of themselves for their account image, and link to their accounts on other online platforms including Linkedin and Twitter. Users build their portfolio of designs over many years. They accumulate reputation by gathering views and likes (engagement) on their images, called shots on the site, and followers on their account. The social network aspect of the site facilitates continued interactions as users see more and more of each others' work. Success on Dribbble has impact outside the site itself, as it can bring significant employment opportunities and influence. The platform has recently added a job board and special recruiter accounts. 

\begin{figure}[t]
	\centering \includegraphics[width=\columnwidth]
	{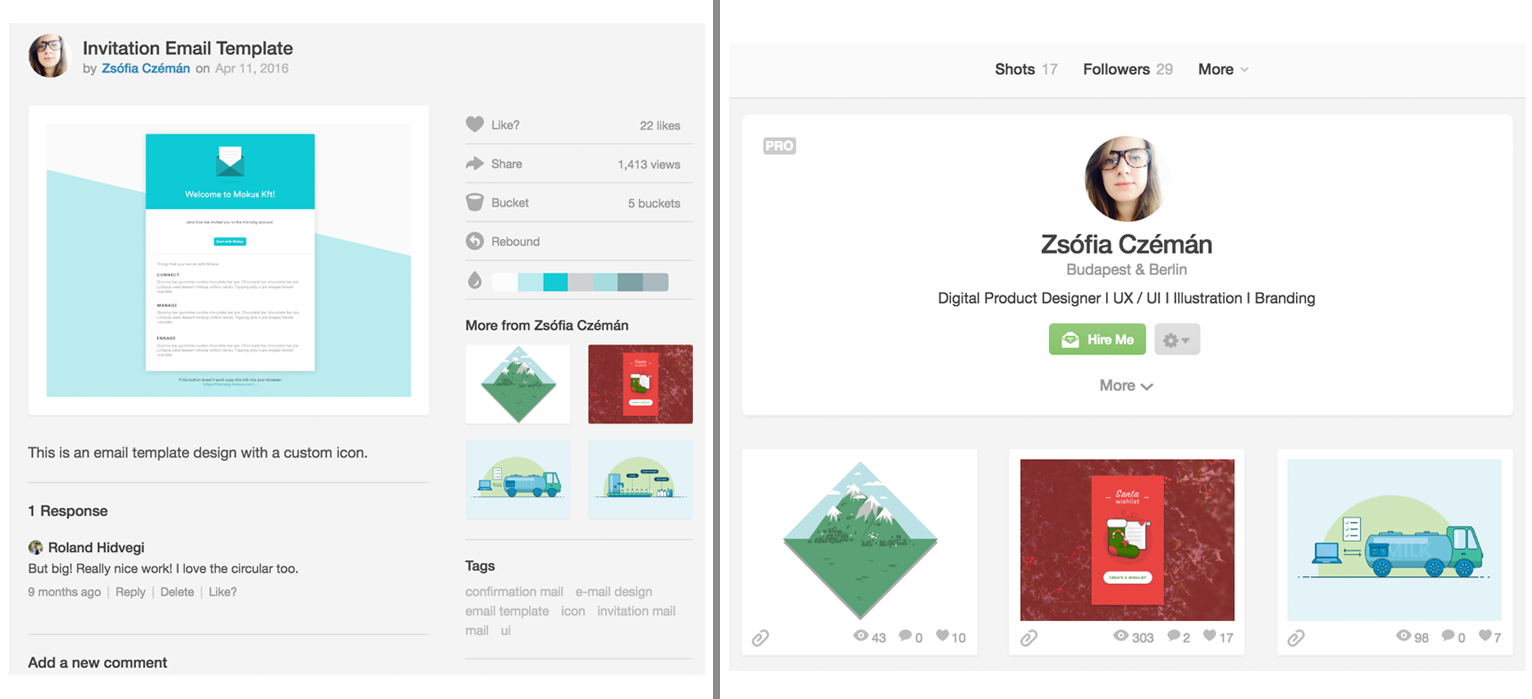}
	\vspace{-1em}
	\caption{Shot (Image) and User Pages on Dribbble.}
	\label{fig:shot_user}
\end{figure}

\subsection{Data Collection}
Our data sample consists of all Dribbble users who were members of a team at the time of the data collection. Typically companies form teams on Dribbble as paid umbrella accounts that users can join. We select this sample in order to gather a comparable set of users who are both active and committed members of the site. We then crawled the profiles of 6,215 users identified as team members. Next, we crawled all 60,406 images made by these users. In subsequent analysis, we discard users making fewer than five images\footnote{Our results are robust to including these users.}. We also discard images posted by the team account with identifiable individual author. We share examples of an image and a user page in Figure~\ref{fig:shot_user}. Data collection took place between September and November 2016 and observed listed rate-limits on the Dribbble API. 

\subsection{Extracted User features}
At the shot level we first record the image itself, the date it was made, and the identity of the author. We also note the tags the author annotated the shot with. Tags are free-form key words that say something about the image. Others can search for images listing specific tags. Tags therefore serve a dual purpose: to describe what the author is doing, and to help others find the image. Each shot has a count of the likes that it received, which can be thought of as the main success measure in the community. 

At the user level we collect the name of the author, whether the author has a ``pro-badge'', and the author's tenure on the platform (in days). A pro-badge is a sign that the user has paid for a premium account, which facilitates job search features on the site and lifts the cap on the number of shots a user can make in a given amount of time. We consider pro-badges as a proxy for buy-in on the platform. At the shot level we calculate how many shots a user has made before to quantify their experience. Finally, we also estimate the gender of each user. Since the profiles do not directly list gender, we infer them from the users' first names using the US baby name data set~\cite{babyname}. For any user with a name not in the database or an ambiguous gender score (i.e. greater than 10\% and less than 90\%) we manually check their self-portrait on Dribbble and on linked social media accounts.

\subsection{Network Features}
Like many other online communities, Dribbble is built on top of a social network. When a user follows another user, the second user's future shots are included in the default newsfeed of the first user and so following a user has bandwidth costs. We collect a list of all following relationships amongst our users and when they were created. These timestamped edges allow us to recreate the social network of our users at the time when an image was submitted . For each image we calculate several network measures quantifying the position of the user at the time of creation.

\begin{itemize}
\item{\textit{In-degree:} How many followers the user has.}
\item{\textit{Out-degree:} How many other users the user follows.}
\item{\textit{Closeness centrality:} One over the average distance of the user from all other nodes~\cite{bavelas1950communication}. This measures how close the user is to the center of network.}
\item{\textit{Constraint:} Burt's measure of the extent to which a user's outgoing connections are redundant~\cite{burt2004structural}}.
\item{\textit{Density:} The ratio of observed ties to possible ties among the users the user follows.}
\end{itemize}

In- and out-degree quantify the simple connectivity of a user. Closeness centrality is a global network measure which increases as the user is closer to the center of the network. Constraint and density of the user measure the cohesiveness of his local social network.

\begin{figure*}[t]
	\centering \includegraphics[width=0.98\textwidth]
	{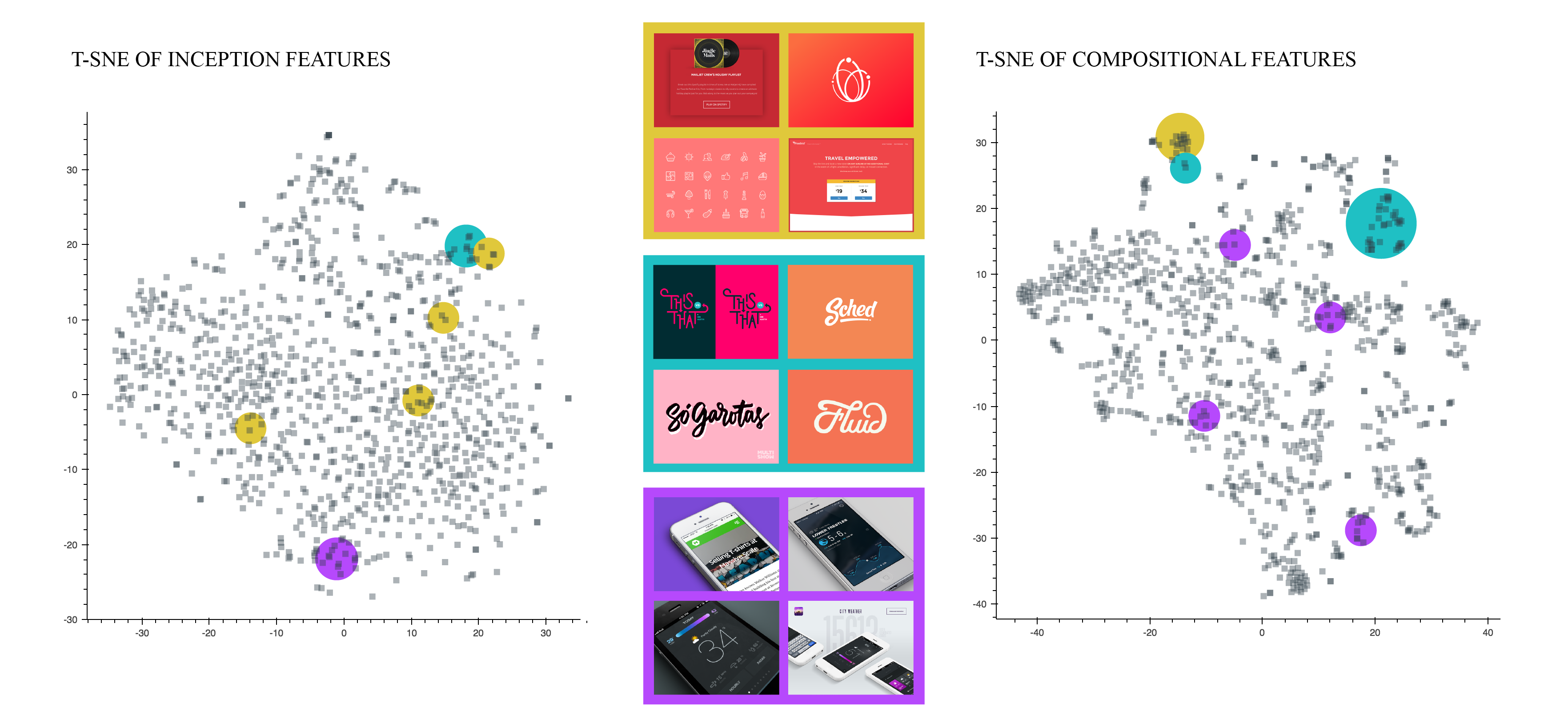}
	\vspace{-1em}
	\caption{Visualizing sample images using t-SNE dimensionality reduction of Inception and compositional features. We highlight three example groups of images. Images from the gold group are close together in the compositional feature space but spread out in the Inception feature space. The teal group is close in both feature spaces, with one exception in the compositional space. Images from the purple group are close in Inception space but scattered in compositional space. The gold group consists of a logo, a collection of icons, a web page design, and an email flier: they are likely clustered in compositional space because of their color. The members of the purple group are all mobile phone screens. Members of the teal group are likely clustered in both spaces because they share both structural and compositional qualities.}
	\label{fig:tsneviz}
\end{figure*}

\section{Extracting Image Features}

In this section we describe two sets of images features upon which we calculate an image's novelty. First we calculate \textit{compositional features}. Then we use a neural network framework to extract a set of unsupervised features. We compare the two feature spaces by projecting them to a low-dimensional space in which similar images are placed closer to one another. We examine what kind of images are similar according to the two feature sets, finding that the compositional features capture color and style while the neural network features capture content. 

\subsection{Compositional Features}
Imitating precisely previous work on the qualitative features of images~\cite{schifanella2015image}, we define 47 compositional features for each image. These features are derived from aesthetic considerations and have proven to have significant predictive power of the beauty or attractiveness of images. Previous work groups the features into the following categories: colors, spatial arrangements, and texture.

Color features include contrast (defined in terms of luminance) and the averages of hue, saturation, and brightness across both the whole image and a subset in its center~\cite{datta2006studying}. We also include three ``emotional'' features which are linear combinations of saturation and brightness: pleasure, arousal, and dominance~\cite{machajdik2010affective}. Binning hue, saturation, and brightness yield \textit{Itten Color Histograms} and taking their standard deviations yields \textit{Itten Color Contrasts} after a careful segmentation. Spatial features include symmetry and salience~\cite{hou2007saliency}, the distribution of which describes how attention-grabbing different regions of the image are. Finally, Haralick's texture features quantify image complexity: entropy, energy, homogeneity, and contrast~\cite{haralick1979statistical}.

\subsection{Neural Network Features}
Feedforward-based neural networks have made tremendous strides in object-in-image classification tasks in recent years. Many such networks have penultimate layers which reduce images input for classification into a feature space for the classification layer. It is possible to extract these features from pre-trained neural networks. We harness one such network: the Inception v3~\cite{szegedy2015going}, originally constructed to optimally classify a large dataset of images into 1000 categories. We acknowledge here that there are many alternative specifications to generate similar sets of features. Passing our images through the network we generate 2048 features that encode highly discriminating facets of the data.

\subsection{Visualizing Image Features}
Before proceeding, we pause to visualize and inspect our data in the two visual feature spaces. We reduce the 47 and 2048 dimensional spaces to two-dimensions using t-SNE, a popular dimensionality reduction method that uses information theoretic methods to minimize distances between data points in the projection as a function of their similarity~\cite{maaten2008visualizing}. In Figure~\ref{fig:tsneviz}, we visualize the 2-D t-SNE projections of a random sample of 200 images a year from 2012 to 2016 using the Inception and compositional features, respectively.

In both projections we observe the clustering of images into groups. The qualitative attributes that define the clustering, however, are quite different. As highlighted in Figure~\ref{fig:tsneviz}, clustering on compositional features is based on color and aesthetic style, as expected. In the projection based on Inception features, however, we observe that images cluster based on their content. In other words, images with highly similar Inception features are likely to represent similar concepts, be they logos, mobile phone interfaces, icons, wire-frames, etc. This is perhaps not surprising given Inception's origin as an object-in-image classification tool. This characterization of the two features sets as describing style and content is important for understanding their novelty.

\section{Novelty measures}
In this section we define a reference novelty based on user annotations or tags of an image by defining the relative surprise of seeing a set of tags on image, compared with the tags that came before. We then define a measure of novelty for our visual feature spaces using Gaussian mixtures and Fisher information.

\subsection{Tag Novelty}

Before calculating novelty using visual features, we create a novelty measure using the tags an author gives an image. Following \cite{sreenivasan2013quantitative}, we calculate the ``surprise'' of each tag of an image. That is, given all the images and their tags posted before the image, we define the probability of observing a tag $t$ as $P(t)$, the proportion of previous images listing that tag. The log of $P(t)$ is our measure of the surprise of a tag. As we are especially interested in completely new tags, we also include the focal image and its tag when we calculate $P(t)$, to avoid taking the log of 0. We then define tag novelty $N_{i}$ of an image $i$ with tags $t_{1},t_{2},\ldots t_{n} \in T_{i}$ as the aggregate the surprise of an image's tags:

$$N_{i}= -\dfrac{1}{|T_{i}|} \sum_{t \in T_{i}}{\log P(t)}$$

In order to make our measure robust to the order of the images, we scale each image's tag novelty by the maximum possible novelty. Namely, if $I$ is the number of images made before image $i$, we normalize the equation above by $-\log(|I|)$.

\subsection{Visual Novelty via Fisher Information}
To study the visual novelty of images, we define a parametric model for images in terms of their position in a given feature space. Given a new image, we consider the distribution of previous images in a feature space and approximate them using Gaussian mixture models. We calculate the likelihood of the focal image relative to these distributions using its Fisher information, an information theoretic measure which we prefer to alternatives such as the Akaike information criterion because of its reparametrization-invariance. Specifically we define novelty as one minus the norm of the Fisher vector of an image over the Gaussian mixture models. This approach is similar in style to a recent method to calculate novelty using a data point's distance to the centroids of a k-means clustering~\cite{redi2014}.

Formally, let be $x \in \mathbb{R}^d$ a finite d-dimensional real representation of an image and a parametric model $p(x | \theta)$ where $\theta$ is the parameter of the density function. If the model is a Gaussian mixture model (GMM) with $N$ Gaussians, the pdf is $p(x|\theta) = \sum_i^N \omega_i g_i(x)$ where the $g_i(x)$ is the density function of the $i$-th Gaussian. The continuously evolving model changes the parameters of the probabilistic model with the emergence of new images in time. We consider two different likelihood measures to apply to the probabilistic model:
\begin{itemize}
\item Akaike's information criterion (AIC)~\cite{akaike1981likelihood}: we measure the AIC per image according the actual state of our generative model.

\item Fisher information: after calculating the Fisher score~\cite{jaakkola1999exploiting} of for each image according to the shape of the model we can measure the similarity of images $x$ and $y$ with the Fisher kernel, as
\begin{equation}
K_{\theta}(x,y) = \nabla_{\theta} \log p(x |\theta)^T F_{\theta}^{-1} \nabla_{\theta} \log p(y |\theta)
\end{equation}
where $F_{\theta}$ is the Fisher information matrix. The gradient of the likelihood indicates how the model may change to fit the actual point, in our case an image. Our choice was driven by the unique invariance properties (e.g. reparametrization invariance) of the Fisher information matrix and the Fisher kernel ~\cite{Cencov1982,campbell1986extended,lebanon2004extended}. Applying Cholesky decomposition, the kernel can be defined as a simple scalar product, as $K_{\theta}(x,y) = \mathcal{G}_\theta(x)^T \mathcal{G}_\theta(y)$
where $\mathcal{G}_\theta(x)=\nabla_{\theta} \log p(x |\theta) F_{\theta}^{-1/2}$ is the normalized Fisher score or the Fisher vector of image $x$. We note that the Fisher vector has dimension $\mathcal{O}(d|\theta|)$.
\end{itemize}

On account of its reparametrization invariance we choose to continue with the Fisher information as our measure of likelihood. Although estimation of the Fisher information matrix is difficult, there are known closed form approximations for both Gaussian mixture models~\cite{perronnin2007fisher} and special classes of Markov random fields~\cite{daroczy2015text}. We suggest two potential definitions of novelty measures based on the Fisher information:

\begin{itemize}
\item Norm of the Fisher Vector over Gaussian Mixture (FVGMM): as the Fisher score highlights how the model parameters should change to best fit the focal image, our first novelty measures the norm of the Fisher vector for each image as 
\begin{equation}
N_{FV}(x) = ||\mathcal{G}_{\theta}(x)||=||\nabla_{\theta}\log p(x|\theta) F_{\theta}^{-1/2}||.
\end{equation}
In case of Gaussian Mixtures the pdf is $p(x|\theta)=\sum_{i}^N \omega_i g_i(x)$ where $\theta$ consists of the mixture weights, mean, and covariance parameters of the Gaussian mixture. In practice we observe that the Fisher score for both compositional and Inception features is very sparse because of the ``peakness'' property of the \textit{membership probability}, defined as the probability that a point is generated from one of the Gaussians. In comparison with~\cite{askin2017makes} this method puts the most weight on the most similar images that came before the focal image.

\item Similarity graph over the Gaussian Mixture (FVMRF): one approach to overcoming the ``peakness'' property while still capturing the temporal distribution is to define a Markov random field following~\cite{daroczy2015text} with the mean of the Gaussian mixture as the sample set. The main idea is to define an undirected random field, which is a graph with $N$ nodes consisting of random variables and sample points, connected to our image as a separate random variable in a star. The probability density function of the new distribution can be factorized over the maximal cliques in the resulting graph. In our case the edges and therefore the pdf are:
\begin{equation}
\label{eq:mrf}
p(x|\alpha, \theta) = \frac{e^{-\sum_i \alpha_i ||x-\mu_i||}}{\int_{x \in \mathcal{X}} e^{-\sum_i \alpha_i ||x-\mu_i||} dx} 
\end{equation}
where $\mu_i$ is the mean vector of the $i$-th Gaussian, $\alpha$ is the relative importance of the cliques. The Fisher vector can be approximated in this context with a simple formula~\cite{daroczy2015text}:

\begin{equation}
\label{eq:mrf_end}
\nonumber
N_{FVMRF}(x) = \{\frac{d_i(x)-\mathbf{E}[d_i(x)]}{Var^{-\frac{1}{2}}(d_i(x))}\}
\end{equation}

where $d_i(x) = ||x-\mu_i||$ and $i \in {1,...,N}$.
\end{itemize}

Given the relatively high complexity of the random field approach, we define novelty using the norm of the Fisher vector\footnote{In applications where the aforementioned peakness issue is more pronounced, we recommend using the random field approach}. As the method returns a similarity score, we subtract one to define visual novelty. For the rest of the paper we refer to this novelty score as \textit{Inception novelty} when it is calculated using Inception features, and \textit{compositional novelty} when it is calculated using compositional features. 

\subsection{Comparison of Novelty Scores and Validation}
\begin{figure}[tp]
	\includegraphics[width=\columnwidth]{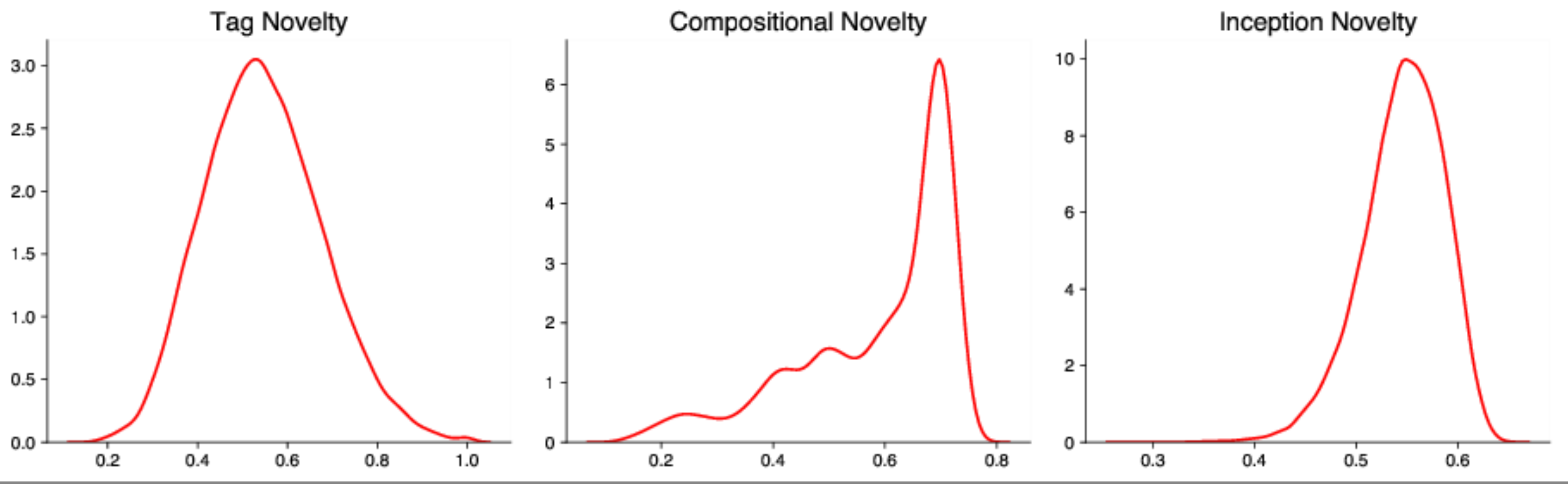}
	\caption{Kernel density estimated distributions of tag, compositional, and Inception novelty.}
	\label{fig:novdists}
\end{figure}

We visualize the distribution of tag, Inception, and compositional novelty scores in Figure~\ref{fig:novdists}. We correlate the two novelties with tag novelty and several user-level features in Table~\ref{corrmat}. We find that both visual novelties are weakly correlated with tag novelty. The correlation is roughly twice as strong for Inception novelty than compositional novelty. This suggests that tags are used to describe images in a conceptional rather than stylistic manner. The two visual novelties are significantly correlated, and, together with tag novelty, are negatively correlated with engagement. We note that the platform's design may explain the trade-off between engagement and tag novelty: users can search for images by tags.

\begin{table}[bp]
\centering
\tabcolsep=0.1cm
\begin{tabular}{rllll}
  \hline
 & (1) & (2) & (3) & (4) \\ 
  \hline
  Tag Novelty (1) \\
  Inception Novelty (2) & 0.123  &  &  &   \\ 
  Compositional Novelty (3) & 0.067 & 0.274 &  &    \\ 
  Likes (Log) (4) & -0.138 & -0.082 & -0.014 &     \\ 
  Views (Log) (5) & -0.138 & -0.114 & -0.058 & 0.927  \\ 
   \hline
\end{tabular}
  \caption{Correlation matrix of novelty and success features.} 
  \label{corrmat} 
\end{table}

\subsubsection{Validation of Visual Novelty}
As discussed, novelty is an ephemeral quality of a cultural product and its measurement implicitly requires comparison, more so than, for example, its beauty. We cannot, for instance, ask someone to evaluate the novelty of a four-year-old mobile phone application layout. In this case success and perceptions of novelty are likely anti-correlated: success breeds familiarity. 

\begin{figure}[t]
	\includegraphics[width=\columnwidth]{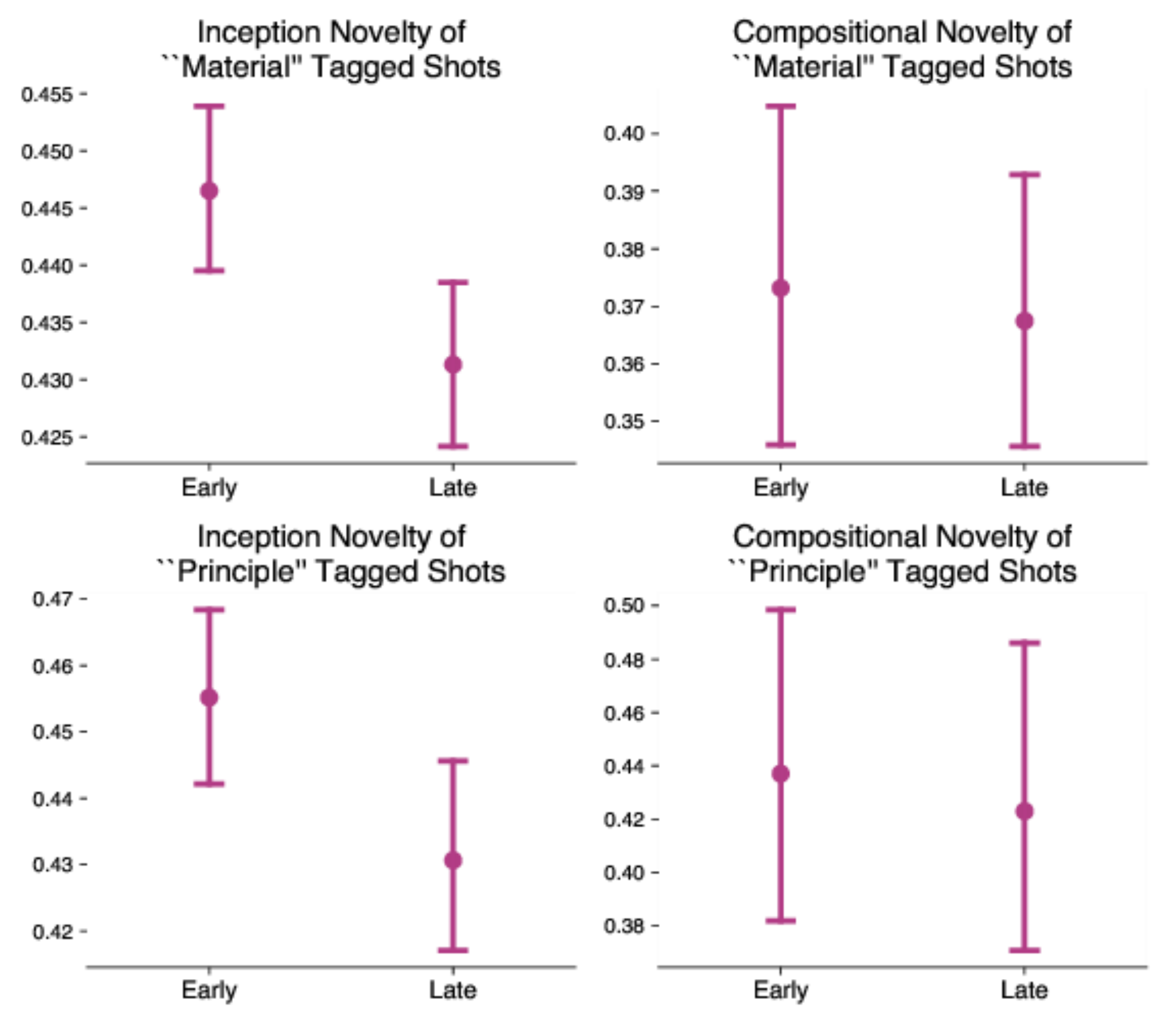}
	\caption{Comparison of visual novelty scores of images with the tags ``material'' and ``principle''. We consider those images in the first 10\% and most recent 10\% of all images created using the tags. We find that Inception novelty is significantly higher for images listing these ``emerging'' tags.}
	\label{fig:matprincnov}
\end{figure}

One approach to validate our measures of visual novelty, besides the correlations with tag novelty noted above, is to identify a population of images which are likely to be covering a new kind of product that emerges in the middle of our dataset. We identify emerging product types by finding tags which are used only after 2013, yet still are among the 200 most used tags. We find two such tags\footnote{Other examples of tag fitting our quantitative criteria are tags used by groups of designers to indicate group membership. Though these tags certainly merit further study, they do not capture the emergence of a new design approach or method} which we can verify as representing truly emerging novelties: ``material'' and ``principle''. 

Material design\footnote{https://material.io/} is a design language or vocabulary created by Google, announced to the public in June 2014. Like other design languages, it has guidelines and principles that shape the design process, resulting in a consistent look with certain qualities. Material design was created especially for use in digital and technological areas. It emphasizes the use of print design best practices together with motion. Material or ``material design'' appears as a tag in 748 images in our dataset.

Principle\footnote{http://principleformac.com/} is a new software design tool for creating interactive and dynamic user interfaces. Released in August 2015, it is a popular tool for designers to prototype UIs. 243 images in our dataset include a ``principle'' tag.

For both tags we compare the distributions of novelty for the first 10\% of images using the tag, with the most recent 10\% of images using the tag. In figure~\ref{fig:matprincnov} we plot the resulting distributions. We find that Inception novelty is significantly higher for the earliest images tagged with ``material'' (Mann-Whitney U = 1897, p<.01) and ``princple'' (Mann-Whitney U = 190, p<.01) compared with the most recent ones. Though the average compositional novelty is higher for the earliest images in both cases, the differences are not statistically significant (resp. U = 2465, p ~ .26; U = 288, p ~ .32).

\section{Novelty, Networks, and Success}

In this section we investigate which users are more likely to create novel images and whether novel images are more or less likely to be successful. We consider both Inception and compositional novelty.

\begin{table*}[t] \centering 
\begin{tabular}{@{\extracolsep{2pt}}lcccccc}
\\[-1.8ex]\hline 
\hline \\[-1.8ex] 
 & \multicolumn{6}{c}{\textit{Dependent variable:}} \\ 
\cline{2-7} 
\\[-1.8ex] & \multicolumn{3}{c}{Inception Novelty} & \multicolumn{3}{c}{Composition Novelty} \\ 
\cmidrule(lr){2-4}
\cmidrule(lr){5-7}
\\[-1.8ex] & (1) & (2) & (3) & (4) & (5) & (6)\\ 
 Days Active (log) & $-$0.015$^{***}$ (0.002) & $-$0.016$^{***}$ (0.002) & $-$0.017$^{***}$ (0.002) & $-$0.000\phantom{$^{***}$} (0.002) & \phantom{$-$}0.000\phantom{$^{***}$} (0.002) & $-$0.000\phantom{$^{***}$} (0.002) \\ 
  nShots Previous &  \phantom{$-$}0.020$^{***}$ (0.007) & \phantom{$-$}0.022$^{***}$ (0.007) &\phantom{$-$} 0.022$^{***}$ (0.007) & \phantom{$-$}0.013$^{*}$\phantom{$^{**}$} (0.007) & \phantom{$-$}0.013$^{*}$\phantom{$^{**}$} (0.007) & \phantom{$-$}0.013$^{*}$\phantom{$^{**}$} (0.007) \\ 
  Male & $-$0.002\phantom{$^{***}$} (0.013) &\phantom{$-$} 0.001\phantom{$^{***}$} (0.013) &\phantom{$-$}0.001\phantom{$^{***}$} (0.013) & $-$0.000\phantom{$^{***}$} (0.013) & $-$0.000 \phantom{$^{***}$}(0.013) & $-$0.000\phantom{$^{***}$} (0.013) \\ 
  Pro & $-$0.036$^{***}$ (0.010) & $-$0.034$^{***}$ (0.010) & $-$0.034$^{***}$ (0.011) & $-$0.034$^{***}$ (0.011) & $-$0.034$^{***}$ (0.011) & $-$0.034$^{***}$ (0.011) \\ 
  In-Degree (log) & \phantom{$-$}0.004 \phantom{$^{***}$} (0.004) & \phantom{$-$}0.002\phantom{$^{***}$} (0.004) &\phantom{$-$} 0.001\phantom{$^{***}$} (0.004) &\phantom{$-$} 0.004\phantom{$^{***}$} (0.004) & \phantom{$-$}0.004\phantom{$^{***}$} (0.004) & \phantom{$-$}0.004 \phantom{$^{***}$}(0.004) \\ 
  Closeness & $-$0.042$^{***}$ (0.008) &  &  & \phantom{$-$}0.001\phantom{$^{***}$} (0.008) &  &  \\ 
  Constraint &  & \phantom{$-$}0.060$^{**}$\phantom{$^{*}$}  (0.025) &  &  &\phantom{$-$} 0.018\phantom{$^{***}$} (0.026) &  \\ 
  Density &  &  &\phantom{$-$} 0.046$^{**}$\phantom{$^{*}$}  (0.023) &  &  & $-$0.005 \phantom{$^{***}$}(0.024) \\ 
  Constant & \phantom{$-$}0.002\phantom{$^{***}$} (0.025) & $-$0.009\phantom{$^{***}$} (0.026) & $-$0.002\phantom{$^{***}$}(0.026)& $-$0.048$^{*}$\phantom{$^{**}$}  (0.026) & $-$0.053$^{**}$\phantom{$^{*}$} (0.026) & $-$0.047$^{*}$\phantom{$^{**}$}(0.026) \\ 
 \hline \\[-1.8ex] 
Observations & 37,799 & 37,799 & 37,799 & 37,799 & 37,799 & 37,799 \\ 
Log Likelihood & $-$25,740.880 & $-$25,749.400 & $-$25,750.350 & $-$25,731.900 & $-$25,730.540 & $-$25,730.860 \\ 
Bayesian Inf. Crit. & 51,576.620 & 51,593.660 & 51,595.570 & 51,558.660 & 51,555.950 & 51,556.580 \\ 
\hline 
\hline \\[-1.8ex] 
\textit{User random effects}& \multicolumn{6}{r}{$^{*}$p$<$0.1; $^{**}$p$<$0.05; $^{***}$p$<$0.01} \\ 
\end{tabular} 
\caption{Predicting novelty with network position.} 
\label{predictingnov} 
\end{table*}

First we use hierarchical linear regression~\cite{gelman2006data}  on data at the image level with user random-effectsand controls to predict novelty. Our aim is understand who makes novel images. Then we predict success using novelty and network position. In both cases we control for gender, the (log) number of shots made previously, the (log) number of days the user has been active on the site at the time of the shot, and whether the user has a paid account. In other words we control for gender, productivity/experience, tenure, and investment into the site.

\subsection{Who makes novel shots?}
We find several significant predictors of Inception novelty, both among our control variables and network variables. Interestingly, the network features we consider do not impact compositional novelty. We summarize these findings in Table~\ref{predictingnov}.

For both compositional and Inception-based measures we find that pro-users are less likely to make novel images. One interpretation is that users who take the site more seriously are more risk-averse and less likely to experiment. Users making more shots in the past make slightly more novel shots. There is mixed evidence that users active for a longer period of time make less novel shots. We detect no gender disparity.

The two novelty measures diverge when we consider the impact of network features. The Inception-based measure of novelty is significantly lower for users closer to the core of the network, and higher for users with cohesive local networks defined by density and constraint. This supports our hypothesis that cohesion facilitates novelty. We find no significant relationship between network position and compositional novelty.

\subsection{When are novel shots successful?}
We now turn to the question of predicting engagement, measured by likes, using novelty. We find that novel shots are generally less successful. We summarize our findings in Table~\ref{novsucnet}. Pro users are more successful, as are those who have many followers. We find that constrained users are less successful. Finally, novel images are in general less successful. 

\begin{table} \centering 
\begin{tabular}{@{\extracolsep{5pt}}lcc} 
\\[-1.8ex]\hline 
\hline \\[-1.8ex] 
 & \multicolumn{2}{c}{\textit{Dependent variable:}} \\ 
\cline{2-3} 
\\[-1.8ex] & \multicolumn{2}{c}{Log Likes} \\ 
\\[-1.8ex] & (1) & (2)\\ 
\hline \\[-1.8ex] 
 Days Active (log) & $-$0.006$^{**}$\phantom{$^{*}$} (0.003) & $-$0.004$^{*}$\phantom{$^{**}$}  (0.003) \\ 
  nShots Previous &  \phantom{$-$}0.086$^{***}$ (0.023) & \phantom{$-$}0.090$^{***}$ (0.023) \\ 
  Male & $-$0.046\phantom{$^{***}$}  (0.044) & $-$0.047\phantom{$^{***}$}  (0.045) \\ 
  Pro &  \phantom{$-$}0.196$^{***}$ (0.037) & \phantom{$-$}0.196$^{***}$ (0.037) \\ 
  In-Degree (log) &  \phantom{$-$}0.371$^{***}$ (0.009) & \phantom{$-$}0.373$^{***}$ (0.009) \\ 
  Out-Degree (log) & $-$0.046$^{***}$ (0.013) & $-$0.046$^{***}$ (0.013) \\ 
  Constraint & $-$0.234$^{***}$ (0.059) & $-$0.233$^{***}$ (0.059) \\ 
  Incep. Nov. & $-$0.108$^{***}$ (0.009) &  \\ 
  Incep. Nov. $\times$ Constraint &  \phantom{$-$}0.084$^{**}$\phantom{$^{*}$}  (0.039) &  \\ 
  Comp. Nov. &  & $-$0.025$^{***}$ (0.010) \\ 
  Comp. Nov. $\times$ Constraint &  & \phantom{$-$}0.017\phantom{$^{***}$}  (0.040) \\ 
  Constant &  \phantom{$-$}2.930$^{***}$ (0.088) & \phantom{$-$}2.908$^{***}$ (0.089) \\ 
 \hline \\[-1.8ex] 
Observations & 37,799 & 37,799 \\ 
Log Likelihood & $-$36,353.290 & $-$36,450.650 \\ 
Bayesian Inf. Crit. & 72,833.060 & 73,027.780 \\ 
\hline 
\hline \\[-1.8ex] 
\textit{User random effects} & \multicolumn{2}{r}{$^{*}$p$<$0.1; $^{**}$p$<$0.05; $^{***}$p$<$0.01} \\ 
\end{tabular} 
  \caption{Predicting success with novelty and network position.} 
  \label{novsucnet} 
\end{table} 
 
\begin{figure}[t]
	\includegraphics[width=\columnwidth]{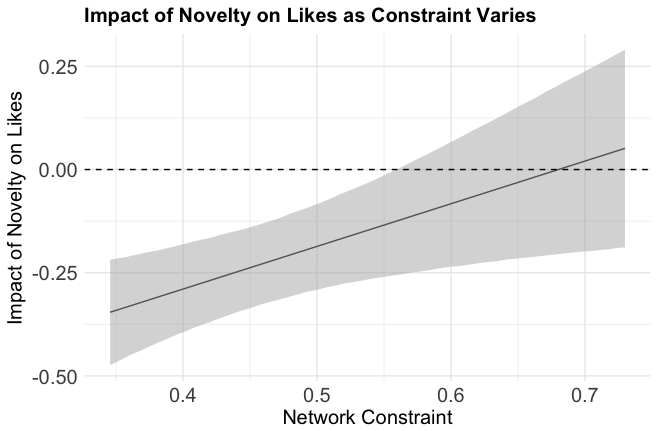}
	\caption{Relationship between success and novelty as constraint varies. Low constraint users have less success when making novel shots. High constraint users have more success with novel shots.}
	\label{noconstraint_interact}
\end{figure}

We find an interesting interaction between constraint and Inception novelty. Namely, users embedded in highly constrained networks making novel images do better than those in unconstrained networks making novel images. To better interpret this finding we visualize this relationship in Figure~\ref{noconstraint_interact}. In other words, the least constrained users have a penalty for novelty while the most constrained users have a bonus for novelty. We also find a significant interaction between inception novelty and closeness centrality: novelty has an increasingly negative relationship with success as a user is more central in the network, but no relationship between local density and either novelty measure.

\begin{figure}
	\includegraphics[width=\columnwidth,scale=1.1]{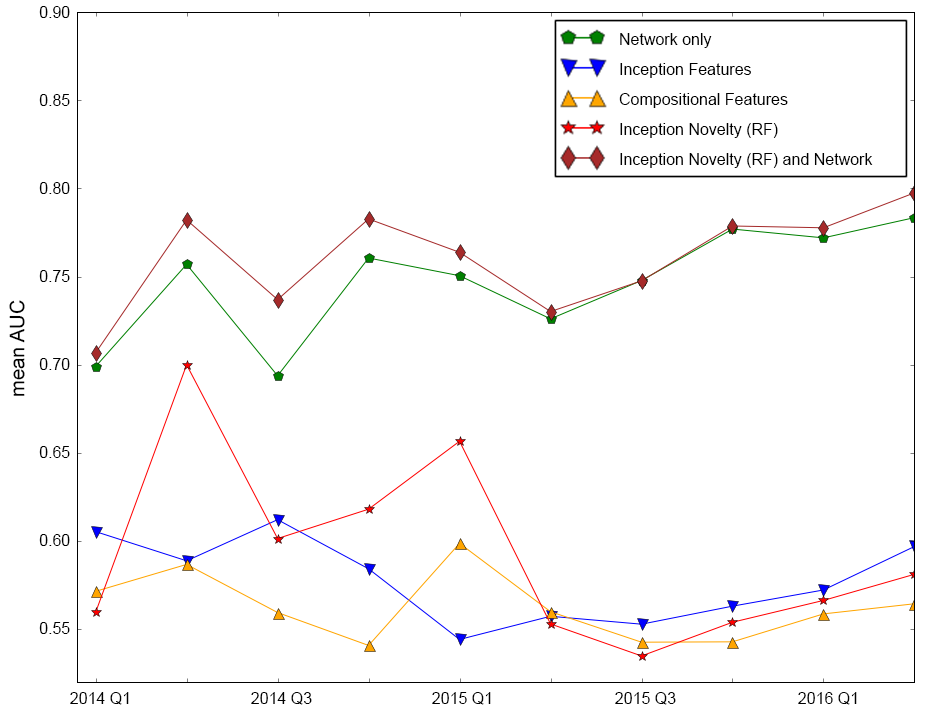}
	\caption{Average AUC of quarter to quarter success prediction. We predict success of the images using gradient boosted trees on the visual image features, novelty scores, and network position of the users. We find that novelty scores extracted image features increase the predictive power of the model including the network features, This suggests that network position does not entirely mediate the relationship between novelty and success.}
	\label{prediction_gbt}
\end{figure}

Finally, using a machine learning framework, we check how well our features can predict success binned into three separate class labels: less than ten likes, between ten and one hundred likes, and more than one hundred likes. As an initialization we used the first year as the first training period and for every consecutive quarter thereafter we consider the previous year. We found that the random field approach to calculating the Fisher vector (FVMRF) was most effective in predicting engagement. Using the area under the receiver operating characteristic curve (AUC), we find that a gradient boosted trees model~\cite{friedman2001greedy} on network and content features has significant predictive power. As we can see in Figure~\ref{prediction_gbt}, even though the network features are the best indicators of success, the content and novelty of the images, encoded using the Inception-based Fisher vectors offer additional predictive power. This suggests that it is possible to use image features to predict success on the site. It is likely possible to do better if features are extracted with the aim of predicting success.

\section{Conclusions}

In this paper we developed, evaluated, and compared measures of novelty of images using data from an online community of digital designers. We first compared different feature sets of images, noting that compositional features like entropy, contrast, and brightness capture qualitatively different facets of an image and features derived from an Inception neural network learning framework capture qualitatively different facets of an image. Specifically, compositional features seem to capture stylistic aspects and while Inception features capture content, in line with their origins.

Next, we created a mathematical framework to compare images with all images that came before in terms of either set of image features. To calculate the novelty of an image, we estimate the distribution of previous shots in the given feature space using a Gaussian mixture model. We then calculate the likelihood of the the image - in other words we quantify how statistically similar the image is to those that came before. We define novelty of an image as one minus this similarity score. 

We find that both novelties calculated from the Inception features and compositional features are significantly correlated with a measure of novelty based on author text annotations or ``tags'' of their images. We also found that Inception novelty was significantly higher for images created in the early stages of an emerging tag compared with images using the same tag later. 

Attempting to understand the profile of a user who makes more novel shots, we turned to the site's social network. Using temporal following data, we related social network position at the time of the creation of an image to its novelty. We found that users with cohesive local networks (quantified by density or Burt's constraint measure) tend to post images with higher Inception novelty.

We also find that users close to the center of the network, in a global sense, make less novel shots. Users with a ``pro-badge'' (paid account) likewise make less novel images. Given the professional atmosphere of the site, including for example its invitation-only participation, the presence of significant players and companies in the field, and the potential for economic opportunities, it seems reasonable that established designers may have reason to make more conventional images. That Dribbble is an online community only compounds the potential costs of creating unsuccessful novelty: though a designer's support system and network of strong ties cannot vastly grow, her audience can scale drastically. The underestimated permanence of online identities makes this asymmetry all the more important when we consider what it means for a designer to take a risk with a distinctive image.

Indeed professional online communities present a dilemma for users in general. Though the feelings of anonymity and distance may facilitate bold experimentation, members of online communities who wish to leverage their investment of time and effort into professional advancement must credibly link their online identities to their real ones. Even users who want to stay anonymous often have a hard time doing so~\cite{horvat2012one}. Once this identification has occurred, the individual must consider that anything they share online is widely broadcast and more consistently recorded and preserved than what they may say or share offline. We claim that as the labor market becomes increasingly digital, online social networks merit closer study.

Turning to the relationship between novelty and success, we find that novelty is related to worse outcomes. We also find that users in highly constrained positions are less successful. On the other hand, the interaction between constraint and novelty is positive: users with cohesive local networks of strong ties making novel images find more success. We argue that these relationships merit further study. Are these embedded designers better positioned to take risks? Can we interpret images with high novelty score, according to our definition, as being risky? The negative relationship between novelty and network centrality raises even more questions.

Our study has several limitations. Given the transient nature of novelty, we have only limited tests of validity for our measures. Given the ubiquity of digital technology, a highly novel digital design from five years ago likely looks highly outdated now. Moreover, the networking behavior of designers on this platform is highly tailored to the situation. For example, users adopting a strategy of aggressive following anticipating reciprocity, may end up in highly dense networks. All at once, Dribbble serves as a social network, professional portfolio, information network, and status hierarchy for the field. Any attempt to infer causal relations between social network structure and the creation of new ideas on this platform must disentangle the complicated layers driving interactions. We also concede that novelty is multi-faceted: no single measure can totally capture such a broad concept. In future work we aim to better understand influence and spreading of novelty.

\section{Acknowledgements}
The authors wish to thank Zs\'ofia Cz\'em\'an, Anna May, and anonymous referees for their helpful suggestions. This research has been funded in part by NSF grant IIS-1514283. D.B was supported by the Momentum Grant of the Hungarian Academy of Sciences (LP2012-19/2012).

\bibliographystyle{ACM-Reference-Format}
\bibliography{websci_novelty.bib} 

\end{document}